# Selective light-induced mass transport in amorphous $As_xSe_{100-x}$ films driven by the composition tuning: effect of temperature on maximum acceleration

M.L.Trunov[1,2]* and P.M. Lytvyn[3],

[1]Institute for Information Recording NAS Ukraine, Kyiv 03113, Ukraine
[2]Uzhgorod National University, Uzhgorod 88000, Ukraine
[3]V. Lashkaryov Institute of Semiconductors NAS Ukraine, Kyiv 03028, Ukraine
E-mail: trunov.m@gmail.com

**Abstract**

Under irradiation of amorphous $As_xSe_{100-x}$ films by band-gap light, it was observed that lateral mass transport of the film material changed the direction of movement in the region ($4 < x < 5$ at%.) of topological structural transition (crossover "from light to dark" to "from dark to light"). We propose a model that qualitatively describes the observed phenomenon. In experimental testing of the model by Kelvin probe force microscopy, for the first time to our knowledge, it was detected that with the increase of temperature (from ambient to glass-transition temperature, $T_g$) photo-induced mass transport in glassy semiconductors occurs with the rates, which significantly exceed those at room temperature.

## 1 INTRODUCTION

Photo-induced mass transport (MT) in thin semiconductor chalcogenide glassy (ChG) films, i.e. lateral movement of the material under band gap light irradiation, observed for the first time in the experiments with light wavelength of 514 nm and intensity about $5 \times 10^2$ W/cm$^2$ in $As_2S_3$ films [1]. Subsequent studies [2] showed that the same effect occurs even at much lower light intensities (50-500 mW/cm$^2$) which are in the same range as the intensities used for holographic gratings recording on ChG thin films. For some film compositions, the amounts of transferred material were comparable with initial film thicknesses or even several times exceeded it (giant MT) [3]. Some unique effects were also detected, e.g. inversion of surface relief [4]), low-temperature MT (at 77 K) [5], giant ripples formation under homogeneous illumination [6] and so on. Although several theoretical approaches and models have been developed to describe the mechanisms of MT numerous details in the process remain unsolved until now [7].

Despite that, MT became attractive for the variety of practical application including hybrid waveguides in the integrated optics [8,9], direct optical recording of surface relief gratings (SRGs) in holography [10-12], for the control the optical response of the ChG materials due to changes in surface morphology in optoelectronics [13] and in the plasmonic nanolithography. In the last case, the localized plasmon-polaritons act as a source of near-field effect producing a nanosized surface relief. *Vice versa*, when the surface relief is known, the mapping of the electric field distribution of surface plasmons is possible [14].

However, time of photo-induced surface relief formation achieved now is a several hours that seriously complicate the recording process.

Earlier [11] we have shown, that the efficiency of light-induced lateral MT has a complicated dependence on As content in $As_xSe_{100-x}$ chalcogenide glasses and found that the direction of MT in the holographic recording in $As_{20}Se_{80}$ ChG films may change when a homogeneous irradiation by orthogonally polarized light applied in addition to the linearly polarized recording beams [15]. Moreover, it was established that the direction of MT in a-Se is opposite to $As_{20}Se_{80}$ composition (from "light to dark" and from dark to light", respectively) [11,15].

The details of compositional kinetics of recording for whole As-Se system were not studied, however. On the other hand, it is evident that specific composition should exists in which film material will change the direction of movement "from light to dark" to "from dark to light". Detection of this composition provide an attractive staring point to model light-induced MT and related effects in As-Se amorphous films.

In this Letter, we present an *in situ* study of the SRG formation in $As_xSe_{100-x}$ amorphous thin films ($0 \leq x \leq 40$ at.%) under holographic exposure by band-gap light with relatively moderate intensity at the surface of the layer in different schemes of recording. We investigate how the polarization of the writing and additional beams influences the type of SRGs forming under long-term exposure for different compositions and demonstrate that direction of MT can be reversed by the variation of the composition of the films in the narrow range of the arsenic concentrations (4-5 at. %). A detailed study of this phenomenon allowed us to propose a new model of MT in amorphous semiconductors induced by absorbed light. Moreover, based on this model, we have found temperature acceleration of lateral MT. This phenomenon allows obtaining giant non-saturated SRGs in very short times, which can be compared with those at two-step method exploiting the phenomenon of photoinduced changes in the dissolution rate of amorphous ChG.

## 2 EXPERIMENT

The studies were performed on approximately 1 μm thick films of $As_xSe_{100-x}$ with the value of $x$ = 0, 2, 3, 4, 5, 8, 12, 15, 20, 30, and 40 at%. The films were obtained by thermal evaporation of bulk ChG on glass substrates in vacuum $10^{-3}$ Pa with relatively low deposition rates (3 - 5 nm/s). After the deposition, the Se-rich films ($0 \leq x \leq 5$ at.%) were kept in the darkness for three months in order to stabilize the structure and to eliminate the relaxation processes. Other films ($8 \leq x \leq 40$ at.%) were annealed near glass-formation temperatures (Tg) of appropriate bulk ChG during one hour. The SRGs were then generated by linearly polarized solid state laser (hereinafter referred as "recording laser") with wavelength $\Lambda$ = 660 nm and intensity within 100-150 mW/cm$^2$ using a standard holographic technique. Interference patterns with the periods from 3 up to 10 μm were projected from the substrate side. The orientation of the polarization of the recording beams was either parallel (*p-p*), or perpendicular (*s-s*) to the grating vector. Additional experiments on the formation of SRGs were carried out using homogeneous backlight of the second laser ("additional laser") with the same wavelength and power density polarized perpendicularly to the polarization of recording laser.

For all the samples, we first studied the formation of SRGs as a function of two polarization combinations, such as *p-p* and *s-s*. After that, we investigated the effect of P or S polarized additional beams. The appropriate recording schemes are shown in Fig. 1 (upper part). To monitor the dynamics of the formation of SRGs we used the evolution of diffraction efficiency (DE: the ratio of intensities of the first diffracted beam to the incident one). DE of the recorded SRGs was measured on a fiber spectrophotometer Optic Ocean using a violet laser ($\Lambda$ =404 nm, I = 10 μW) and was taken proportional to the intensity variation of the first diffraction peak in reflection mode. The profiles of the SRGs were investigated by a NanoScope III atomic force microscopy (AFM) (Bruker, Inc.). Some additional AFM experiments exploiting Kelvin probe method was performed to an explanation the obtained results and will be presented in the Section III.

## 3 RESULTS AND DISCUSSION

We start with two compositions that were select as basic materials: amorphous selenium (a-Se) and $As_{20}Se_{80}$. Elemental a-Se is a model glass-former in chalcogenide science and stable SRG can be realized in this simplest but versatile chalcogenide material [16]. On the other hand, $As_{20}Se_{80}$ glass is one of the most efficient materials for relief recording among the large number of Se-based ChG [17].

The results of the evolution of DE of the recorded SRGs for studied $As_xSe_{100-x}$ films with $x$=0 and 20, respectively, are illustrated in Fig.1 a,b as a function of time for three recording schemes (Fig.1, upper part).

In the case of amorphous selenium (a-Se) and *p-p* recording scheme (curve 1 in Fig1 a), a sudden turning on the recording laser (as shown by arrow "recording light On") results in an instantaneous formation of DE that reaches the saturation rather fast (within a few minutes). With further light exposure, DE decreases very slowly and practically disappears but further, we observe a linear growth of DE without the explicit saturation under prolonged exposure during 4000 s.

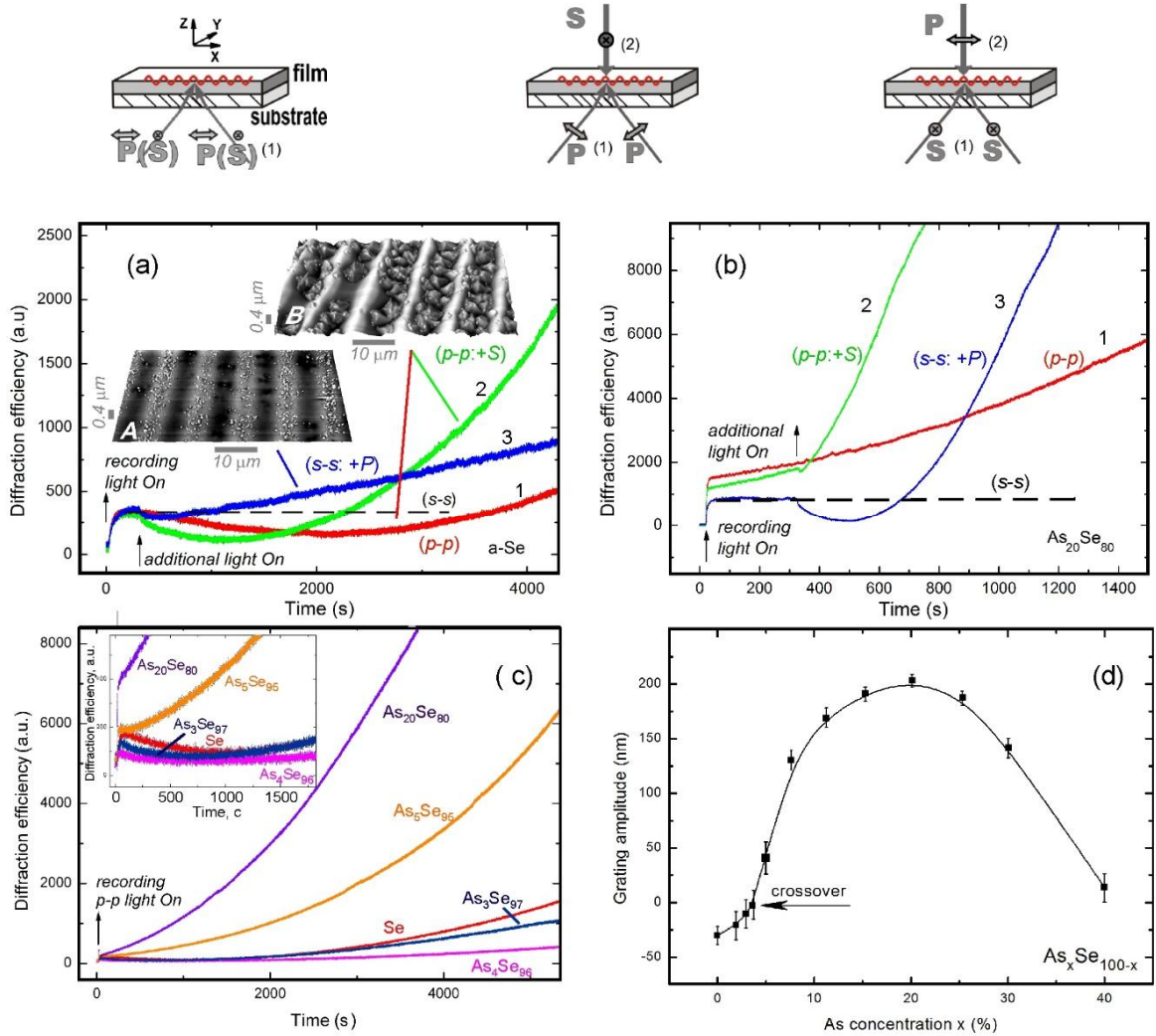

Fig. 1. Variation of DE during SRG formation in a-Se (a) and $As_{20}Se_{80}$ (b) films for different recording schemes (see the sketches in the top images). Polarizations of recording laser (1) and an additional one (2) are shown in the sketches and in the Figure. Arrows show times when recording light (*p-p* or *s-s*) and additional light (P or S) were switched on. DE for *s-s* recording scheme is shown by dotted line. Insets A and B in (a) show AFM patterns of SRG obtained for aged Se films for different recording schemes (see text for details); (c) - Variation of DE during SRGs formation in $As_xSe_{100-x}$ films for *p-p* recording scheme. The concentrations are shown directly in the figure. In the inset: the initial stage of SRGs formation for selected compositions; (d) - the concentration dependence of SRG amplitudes obtained after recording during 1 hour with the laser power P=150 mW/cm$^2$. The arrow shows crossover concentration ($As_{4.5}Se_{95.5}$) at which the direction of MT is changed. The sign of the amplitude is taken as positive for MT direction from "dark to light".

In the *s-s* recording configuration (dashed line in Fig 1 a), the results obtained within a few start minutes of irradiation showed no appreciable changes in the DE kinetics in comparison with the *p-p* scheme (instantaneous DE formation that reaches the saturation very fast), however, the linear growth part of the corresponding DE with continued exposure is absent.

The additional *P* - or *S*- polarized irradiation is expected to affect the DE kinetics (in line with the results of previous studies [15]). In fact, after the additional laser polarized in a direction perpendicular to the recording one was turned on, the DE behavior of the film depended on the recording scheme.

In the (*p-p:*+S) case, when the S-polarized beam was added to *p-p* polarized recording beams (curve 2 in Fig 1 a, arrow "additional light ON") such complicated illumination results in short-term decrease and subsequent rather fast approximately linear increase of DE, respectively. In contrast to the above results for *p-p* scheme of recording, we found that the rate of both processes (DE erasing and linear growing) increases essentially.

In the second case (*s-s*: + P), when the *s*-polarized recording beams and the P polarized light simultaneous illumination were used (see the part after arrow "additional light ON") we observe a linear growth of DE (Fig 1 a, curve 3).

Figure 1 b shows a typical evolution of the DE of the recorded SRGs in $As_{20}Se_{80}$ films in the same different recording configuration. In *p-p* scheme of recording, a sudden turning on of the excitation light (as shown by arrow "recording light On") causes an instantaneous DE formation which increases linearly with further light exposure (Fig. 1 b, curve 1). This part (linear growth) of the corresponding DE in the *s-s* recording configuration (dashed line in Fig 1 b) is absent.

In the scheme with additional light (*p-p*:+S), when the S-polarized illumination was added to *p-p* polarized recording beams (as shown by arrow "additional light ON") the DE growth very fast (Fig 1 b, curve 2) with the rate that exceeds the corresponding rate for *p-p* scheme on one order magnitude.

The opposite situation was observed when recording in the *s*-polarized beams and the simultaneous illumination of P polarized light (*s-s*: + P). Continued exposure with additional light (arrow "additional light ON") almost completely erased the initial DE which appeared when recording *s-s* beams was applied and caused the growth of new DE that linearly increased in time (Fig 1 b, curve 3) with the same rate as in case of *p-p*:+S scheme.

To describe the kinetics of the SRG formation presented above we propose a possible structural model based on photoinduced volume changes [18] and the photofluidity [19] (as a result of photoplastic effect [20, 21]: illumination induced, athermal decrease of the viscosity). We suggest that these two effects are involved simultaneously during the SRG generation and the contribution of each one depends upon the film composition and the polarization scheme of recording. The photoinduced volume change - common to all compositions of As-Se system, albeit with different magnitude [22] takes place at short time scales up to 200 s and as a result the different film density in the bright and dark zones of the interference field leads to the periodic density modulation. This gradient is reflected in the periodic volume change, which causes relatively small SRGs irrespective to the polarization state of recording beams (the scalar effect). The photoinduced lateral MT prevails at longer time scales, i.e., over the extended time of laser irradiation and exhibits vectorial character in the case where the light polarization of the recording or additional beams has a component along the light intensity gradient (see Fig. 1 a, b linear parts of curves 1–3).

In view of the results obtained by irradiating films of a-Se as with holographic recording [16], and at a single beam illumination [23] can be argued that in the recording schemes *p-p* and *p-p*: + S material in MT regime moves "from light" and the schemes *s-s*: + P - "to light", i.e. to the dark and light portions of the interference pattern, respectively.

Direct confirmation of such behavior of the material during holographic recording was obtained in the study of MT in a-Se films, aged in natural conditions over one year. It is known that during prolonged physical aging of a-Se films one can observe their spontaneous crystallization [24], which can be significantly accelerated by additional irradiation by absorbed light [25]. We observed that in the aged but not yet crystallized a-Se films, holographic recording initiates MT, which is accompanied by precipitation of Se crystals, which are located either in the peaks or in the valleys of SRGs, depending on the recording scheme. The *s-s*: +P recording scheme results in precipitation in the peaks of SRG (see inset A in Fig.1a), whereas the *p-p* and *p-p*: +S schemes result in the precipitation in the valleys (inset B in Fig. 1a). Since photo-induced crystallization in Se films can only take place in bright bands of the interference pattern, the variation of MT direction under different schemes of SRG recording (as described above) is completely confirmed.

Since the volume grating is formed in $As_{20}Se_{80}$ by the similar photo-induced expansion of the material in the bright areas of the interference pattern as in a-Se [23], the curves 1-3 (Fig. 1 b) show that MT in $As_{20}Se_{80}$ occurs in the direction opposite to that in a-Se. Material moves to the interference maxima in recording schemes *p-p* and *p-p*:+S (Figure 1 b, curves 1 and 2), and to interference minima for *s-s*:+P scheme (Figure 1 b, curve 3). In the *s-s* recording scheme, we observe only a volume grating formation, like in a-Se (dotted lines in Fig. 1 a, b). We note also that a similar experiment with thin (200 μm) a-Se and $As_{20}Se_{80}$ plates has led to similar results.

Studies of the concentration dependence of MT kinetics in the films $As_xSe_{100-x}$ (Fig. 1 c) using *p-p* recording scheme revealed that the change in direction of MT (crossover) occurs in compositions close to $As_{4.5}Se_{95.5}$, i.e. topological structural transition (5 at.% ≤ x≤4) [26], as it is detailed in the inset in Fig. 1 c. The maximum amplitude of SRG was obtained for the composition $As_{20}Se_{80}$ (Fig. 1 d). The results of measurement with additional illumination are qualitatively similar to those shown in Fig.1 c and not presented here.

Crossover for MT, which we observe in the As-Se system for compositions close to $As_5Se_{95}$, and the lack of SRGs in the films of stoichiometric composition $As_{40}Se_{60}$ (Fig. 1d), can not be explained within the framework of the currently known models based on cooperation so-called "dipolar defects" with polarized light [1], the volume photodiffusion these defects [27,28], by polarizability change under illumination [29] or the viscous flow of the material (light-induced fluidization takes place, viscosity lowers to $10^{12}$ Poise [Ref. 21]) under light pressure and gradient of surface curvature [30]. It is also obvious that the existing methods of structural studies (e.g. micro-Raman studies) do not reveal significant differences between $As_4Se_{96}$ and $As_5Se_{95}$ compounds [31].

Previously [14], we proposed a model based on a driving force, which appears as a consequence of the difference in mobilities of electrons and holes generated by the absorbed light. This difference results in the formation of lateral steady state electric field (lateral photo-Dember effect) which controls kinetics and direction of MT. Later, the appearance of such electric field under homogeneous illumination of ChG films was detected [14, 23] by scanning Kelvin probe force microscopy, which is a unique technique for direct measurements of surface topography and mapping of corresponding surface potential (SP).

Based on the data presented above, we suggest the following mechanism for light induced MT in amorphous chalcogenides. The absorption of band gap photons causes generation of radiation defects and excess electron-hole pairs localized near these defects that lead the formation of positively or negatively charged defects. Due to the lateral steady electric field, there is a lateral diffusion flux of the charged defects. The direction of flow, and hence the direction of the mass transport is determined by the sign of the charged defects and by type of media where the direction of the electric field depends on the carrier mobilities.

The role of polarization of the recording and the additional lasers is also obvious: the MT is initiated only when the electric vector of the recording or additional light is parallel to the SRG vector (i.e. perpendicular to the peaks and valleys).

It is known that the mobility of electrons in a-Se ($6 \times 10^{-3}$ $cm^2/V^{-1}s^{-1}$) is significantly smaller than the mobility of holes ($1.5 \times 10^{-1}$ $cm^2/V^{-1}s^{-1}$), however under illumination at ambient temperature the material shows n-type conductivity with the electron mobility near 0.32 $cm^2/V^{-1}s^{-1}$ (the mobility for holes for such a case was estimated to be very close to the drift mobility in darkness [32]). With small additions of arsenic (3-4 at.%), hole mobility decreases sharply, while the electron mobility gradually decreases, due, apparently, with the emergence of new groups of localized states, controlling the hole mobility [32]. One of these reasons, in terms of changes in the structure may be topological transition observed in $As_xSe_{100-x}$ system at x ~ 4, associated with a change in dimension *D* of glass network (transition from chain-cyclic, ring structure ($D \leq 1$) to chain structure ($D \geq 1$) [31]. A sharp change in the mobility of charge carriers near x ~ 4 can alter the direction of the electric field and, as a consequence, a change in the direction of MT, which reaches maximum rate for composition $As_{20}Se_{80}$. Indeed, electron mobility decreases with the increase of As concentration to 17 – 20 %.

In the films $As_{40}Se_{60}$, due to the extremely low values of electron mobility ($10^{-7}$ -$10^{-8}$ $cm^2/V^{-1}$ $s^{-1}$), the SP, which initiates MT, is practically absent that probably leads the absence of SRGs during irradiation (Fig. 1, d).

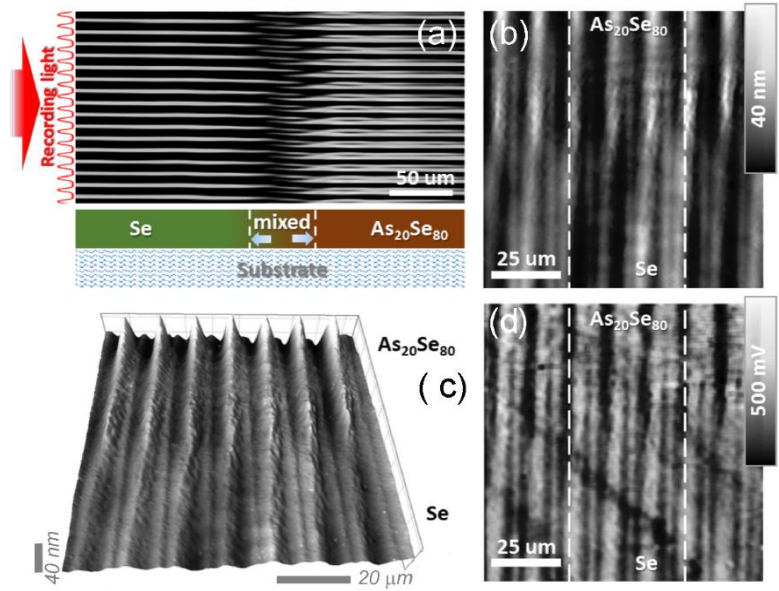

Fig. 2. (a) - The optical image of relative positions of peaks and valleys in SRGs of 10 μm period obtained due to mass transport in the p-p recording scheme in a contact zone between a-Se (left) and $As_{20}Se_{80}$ (right). The arrow "Recording light" shows positions of interference maxima; (b) – AFM image of relative positions of peaks in SRG and (d) surface potential maxima in the contact zone for a-Se and $As_{20}Se_{80}$; (c) - 3D AFM image of the surface relief near the contact.

For a direct confirmation of the proposed model, the following experiments were fulfilled. $As_xSe_{100-x}$ films (for various x indicated above) were deposited on the substrate through a rotatable mask, which allowed to obtain two materials adjacent to each other with the overlap not wider than 50 μm. The a-Se film was selected as the reference material. Irradiation of the contact zone of the films in *p-p* recording scheme (with the spot diameter about 600 μm) allowed comparison relative positions and amplitudes of SRGs obtained for two selected compositions. After that, we tried to observe the distribution of the surface potential (SP) near the contact of both SRGs and compare it with the geometry of forming SRG. For these measurements, we used two-pass Kelvin probe force microscopy technique with the frequency-modulated signal detection [33]. The measurements were performed under ambient conditions using special probes coated by PtIr film. Spatial resolution was better than 50 nm with the sensitivity of few mV.

Fig. 2 show the results obtained for a-Se and $As_{20}Se_{80}$ films. It is seen that different materials show maxima and minima of SRGs shifted for half period (Fig 2 a- c), i.e. direction of mass transport was opposite in both films ("from light to dark" in a-Se where SRG peaks are located at minima of light intensity, and "from dark to light" in $As_{20}Se_{80}$). The results for other compositions (not presented here) show, that the films containing As with x < 4% behave like to pure a-Se (direction of mass transport "from light," and compositions containing As more than 5% - like film $As_{20}Se_{80}$ (direction of mass transport "to light").

However, the SP peaks, unlike SRG ridges, are in phase: maximum SP on the a-Se film coincides with SRG peaks, whereas in the film $As_{20}Se_{80}$ maximum SP coincides with the valleys of SRG [Fig. 2 (d)]. Note that transient case appears near crossover point. Fig. 3 shows the SRGs obtained by *p-p* scheme of recording for the compositions near to point of crossover where the topological transition takes place, namely, $As_3Se_{97}$ (Fig. 3 a) and $As_5Se_{95}$ (Fig. 3 c) and the appropriate SP mapping (Fig.3 b and d, respectively). The corresponding shift of the maximum SP to the valleys of SRG for $As_5Se_{95}$ film is clearly visible, while opposite placement takes place in case of $As_3Se_{97}$ film.

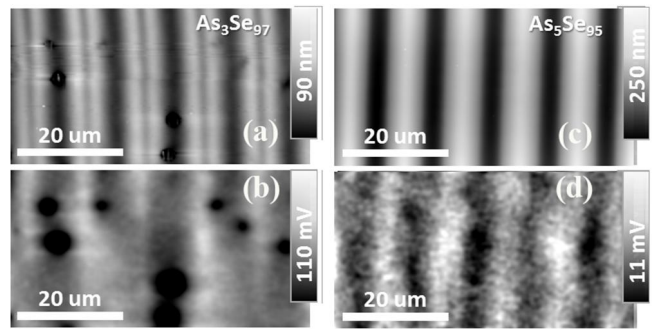

Fig 3 AFM images of SRGs in the compositions near the crossover point: (a)-$As_3Se_{97}$; (c)-$As_5Se_{95}$ and appropriated distribution of SP (b, d).

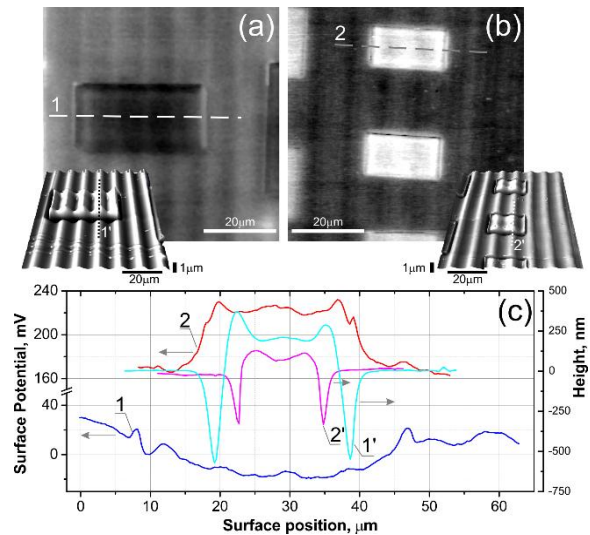

Fig. 4. Maps of SP on SRGs formed with *p-p* recording scheme in a-Se (a) and $As_{20}Se_{80}$ (b) films under irradiation by an electron beam in SEM (rectangles). 3D images of corresponding SRGs are shown in the insets. SPs and SRGs profiles, respectively, along dotted lines 1 (Se), 2 ($As_{20}Se_{80}$) and 1'(Se), 2'($As_{20}Se_{80}$) are shown in (c). Accelerated voltage was 20 keV, current - 7 nA, exposure time - 300 s. The maps were recorded 30 days after SRGs formation under light exposure.

Next, the SP distributions in a-Se and $As_{20}Se_{80}$ films obtained after additional irradiation of SRGs by an electron beam in a scanning electron microscope, SEM (Fig, 4). Original experimental details are not sufficient here and given in [34].

It is seen that SPs in the e-beam irradiated areas of SRGs in these two materials (rectangles in Figs. 4 a and 4 b) differ significantly from the basic matrices. SP in a-Se falls for 40 mV, whereas in $As_{20}Se_{80}$ film SP arises for 60 mV (Fig. 4 c, curves 1 and 2). Both the material move to the irradiated region to form a pedestal (pillow), bordered by a deep depression (see inserts and profiles 1' and 2' in Fig. 4).

It is obvious that the sign inversion of the SP in the e-beam irradiated surface areas indicates that the irradiated a-Se film has *n*-type conductivity, whereas $As_{20}Se_{80}$ film has, respectively, *p*-type. Assuming that the same process also occurs under optical illumination, the direction of MT should be the opposite for these materials.

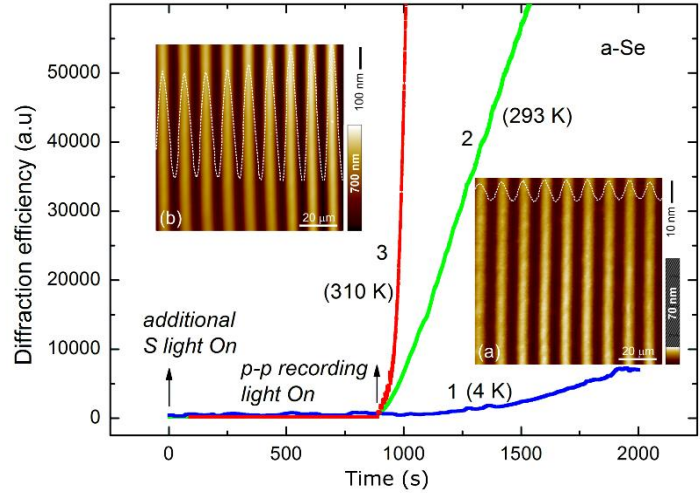

Fig. 5. DE variation during the formation of SRGs in a-Se film at various temperatures (curves 1-3). Temperatures and polarizations of recording and additional lasers are indicated in the figure. The arrows show the moments of switching the light on. Insets: AFM image of the surface relief, obtained at 4 K (a) and 310 K (b). To reduce the influence of cryostat vibrations during recording, the SRGs period was extended to 10 microns.

Since the process of MT under optical and electronic excitation has the general nature [35], the presented results can be described in a unified model. Based on the above approach, the development of such a model requires detailed *in situ* studies of the SP distribution during the SRG formation in a holographic exposure for all three recording schemes [see Fig.1 (a) - (c)]. The results of such studies will be presented elsewhere.

In spite of that, the validity of the above approach can be confirmed under the following assumption.

One can assume that the change in charge carrier mobility, for example, due to significant changes in the recording temperature upon irradiation will lead to a significant change of Dember field, and therefore a change in the MT rate. To test this hypothesis, SRG formation was performed in a-Se films using *p-p*: +S recording scheme at three temperatures: 4 K, room temperature, and a temperature close to the softening temperature $T_g$ (310 K). The result is shown in Fig. 7. Since cooling of the As-Se films leads to a shift of the absorption edge to shorter wavelengths [36], before turning the recording *p*-polarized laser on, a-Se film was pre-irradiated by S-polarized additional laser to saturation of photodarkening.

It also eliminates the appearance of volume gratings due to the photoinduced expansion of film material in the whole zone where SRG will be created. We have detected a SRG of 10 nm amplitude at 4 K [Fig. 5, curve 1 and AFM image (inset a)]. It should be noted that the low-temperature MT so far was observed in Se and $As_{20}Se_{80}$ films cooled only to 77 K [5, 37].

Recording at room temperature resulted in a significant acceleration of the process compared to that at 4 K (Fig. 5, curve 2), while recording at temperatures close to $T_g$, allowed receiving the SRG with amplitude of the order of the film thickness for a few minutes [Fig. 5, curve 3 and AFM image (inset b)]. After recording, the SRG remained stable for a long time after turning the recording laser off and began to erase only at temperatures close to $T_g$ or even well above $T_g$. It is essential that high-temperature MT remained polarization-dependent, i.e. SRG was absent in the *s-s* polarized recording scheme that coincided with the results obtained previously for ChG films at room temperature [11,16].

Within the above model, the results of this experiment can be explained by a significant increase in charge carrier mobility in the a-Se with temperature [38]. On the other hand, taking into account an approach outlined in [28], the result can be caused by temperature dependence of the photo-induced diffusion coefficient.

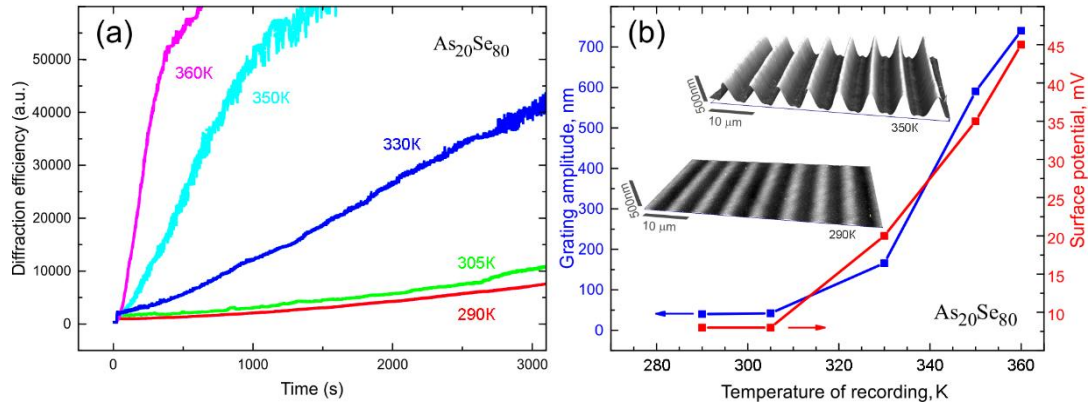

Fig. 6. (a) DE variation during the formation of SRGs in $As_{20}Se_{80}$ film at various temperatures for *p-p* recording scheme. Temperatures of recording are indicated in the figure; (b) SRGs height in $As_{20}Se_{80}$ layers at different temperatures after recording during 15 min with the laser power P=150 mW/cm$^2$ (left) and appropriate them SP (right). Insets: AFM images of the SRGs for two selected temperatures of recording.

Preliminary experiments on high-temperature (i.e. at temperatures close to $T_g$) SRG recording on ChG films of other compositions in As-Se system, in particular, in $As_{20}Se_{80}$ films, showed similar results (see Fig.6 a and insets in Fig. 6 b). Moreover, we have detected that increasing of amplitude in SRG is accompanied by increasing of SP (Fig. 6 b) and this is an additional argument in favor of the proposed model.

**Conclusion**

In conclusion, we characterized the overall dynamics of light induced MT in $As_xSe_{100-x}$ amorphous films over a several main holographic schemes of recording with additional illumination by band-gap light polarized orthogonally to the polarization of recording beams. From the diffraction efficiencies vs time dependences of obtained SRGs we found that crossover occurs in the region of topological structural transition (4 < $x$ < 5 at%.), i.e. the amorphous material change the direction of movement from "light to dark" to "from dark to light". A model that qualitatively describes the observed phenomenon was given that predicts high-temperature photoinduced MT in amorphous chalcogenides with the rates, which significantly exceed those at room temperature. The model was confirmed by direct experiment: a high-temperature initiated giant increasing of MT in some compositions have been detected. Correlating with SP measurements provided by scanning Kelvin probe force microscopy, the increasing of MT is attributed to the lateral photo-Dember effect driven by non-equilibrium carrier dynamics when exciting the film with band-gap irradiation at the temperature near Tg. These findings not only demonstrate a new possible way for fast direct surface patterning, but also reveal the important role of SP dynamics of photomechanical response in amorphous chalcogenides.

**Acknowledgments**

One of the authors (M.L.T.) would like to thank Dr. K. Vad and Dr.V. Takáts (Section of Electron Spectroscopy and Materials Science, Institute for Nuclear Research, Hungarian Academy of Sciences, Debrecen, Hungary) for assistance in the low-temperature measurements (Fig.4, curve 1), and Dr. Cs.

Cherháti and Dr. R. Bohdan (Institute of Physics, University of Debrecen, Hungary), respectively, for the electron irradiation on the SEM (Fig. 4) and for AFM images shown in Fig. 5.